# Automated PMC-based Power Modeling Methodology for Modern Mobile GPUs

Pranab Dash
Purdue University
West Lafayette, USA

Y. Charlie Hu
Purdue University
West Lafayette, USA

Abhilash Jindal
IIT Delhi
Delhi, India

## ABSTRACT

The rise of machine learning workload on smartphones has propelled GPUs into one of the most power-hungry components of modern smartphones and elevates the need for optimizing the GPU power draw by mobile apps. Optimizing the power consumption of mobile GPUs in turn requires accurate estimation of their power draw during app execution.

In this paper, we observe that the prior-art, utilization-frequency based GPU models cannot capture the diverse micro-architectural usage of modern mobile GPUs. We show that these models suffer poor modelling accuracy under diverse GPU workload, and study whether performance monitoring counter (PMC)-based models recently proposed for desktop/server GPUs can be applied to accurately model mobile GPU power.

Our study shows that the PMCs that come with dominating mobile GPUs used in modern smartphones are sufficient to model mobile GPU power, but exhibit multicollinearity if used altogether. We present APGPM, the mobile GPU power modeling methodology that automatically selects an optimal set of PMCs that maximizes the GPU power model accuracy. Evaluation on two representative mobile GPUs shows that APGPM-generated GPU power models reduce the MAPE modeling error of prior-art by 1.95× to 2.66× (*i.e.*, by 11.3% to 15.4%) while using only 4.66% to 20.41% of the total number of available PMCs.

## 1 INTRODUCTION

Optimizing the battery drain of mobile apps helps to extend the mobile device battery life which is critical to enhancing the mobile experience of smartphone users. It requires optimizing the battery drain of all power-hungry device components of modern smartphones, including CPU, GPU, display, Wi-Fi/LTE/5G, GPS, and hardware decoder. Optimizing the power consumption of these power-hungry phone components in turn requires accurate estimation of their power draw during app execution.

The past few years have witnessed a significant increase in incorporating machine learning (ML) in mobile applications with examples ranging from computer-vision-based apps, recommendation-based apps such as streaming apps and social networking apps, Augmented Reality/Mixed Reality, cloud gaming, and language-processing apps such as Siri. In such apps, the mobile GPU runs ML algorithms such as Deep Neural Network (DNN) models in addition to frame rendering. Such heavy load on the GPU significantly increases the utilization of the GPU but more importantly makes the GPU one of the most power-hungry components of modern smartphones.

In this work, we study how to accurately estimate mobile GPU power draw during app execution. While modern smartphones provide an in-built power sensor, it can only measure the total phone power draw and often at coarse-granularity (e.g., about 50 ms on Pixel 4). Similarly, using an external power meter could only measure the total power draw of the phone. As such, statistical power modeling has been widely used to estimate the GPU power draw based on input features that can be easily collected via the operating system. We note that a very limited number of modern smartphones come with manufacturer-specific power rails which are supposed to provide component-wise current draw, *e.g.*, Pixel 7 comes with 86 power rails. Due to their limited availability, power rails cannot be used as a general methodology for GPU power estimation for smartphones. Further, due to the lack of documentation, it is difficult to ascertain how to combine a specific subset of the numerous power rails in a phone to estimate the GPU power.

State-of-the-art mobile GPU power modeling uses utilization-frequency based models (*e.g.*, [10, 25]) which use the GPU utilization and GPU frequency as the input features. However, our measurement study of such utilization-based model using two representative mobile GPUs, Qualcomm's Adreno 640 (on Pixel 4) and ARM's Mali G-710 GPU (on Pixel 7), shows that the widely used utilization-frequency model could not accurately estimate the GPU power across different workloads, as the model cannot capture the diverse microarchitecture-level usage such as instructions and memory transactions breakdown under different workloads.

In contrast, for desktop/server GPUs, researchers observed that modern GPUs come with a comprehensive set of hardware Performance Monitoring Counters (PMCs) that gather various statistics about micro-architectural events in the GPU and memory system during runtime [16], and proposed PMC-based power models that use manually chosen PMCs as input features [21]. PMC-based power models were shown to outperform utilization-based power models. However, manual selection of PMCs suffers several limitations: (1) the selected PMCs may not be optimal, and the resulting power model may provide sub-optimal model accuracy; (2) finding equivalent PMCs for a new architecture may not be easy or possible.

***Contributions.*** In this paper, we study an important research question: *can PMC-based power modeling be applied to mobile GPUs, and if so, what is the right methodology?* Our study makes the following contributions.

First, since mobile devices are power-constrained, the mobile GPU manufacturers may be highly judicious in adding hardware PMCs to the architecture as they incur extra overheads such as silicon die area and design complexity as they need to be read out by the CPU. We study whether the PMCs available on mobile GPUs are comprehensive enough in capturing all architectural components of the GPU and hence the overall power draw of the whole GPU. Our measurement study using multivariable linear regression on the PMCs on the two representative mobile GPUs under diverse workloads shows that the PMCs on modern mobile GPUs can indeed be used for accurate GPU power modeling.

Second, since including inter-dependent variables in linear regression-based modeling can result in lower modeling accuracy and hence be counter-productive, we study whether the PMCs in mobile GPUs are inter-dependent and if they need to be filtered when used in power modeling. Our measurement study shows that multicollinearity indeed occurs when using all the PMCs in modeling the mobile GPU power and we uncover two reasons why this happens: some PMCs are related to each other by conversion, and some PMCs are related by derivation. First, PMCs can be converted from each other, *e.g.,* in each memory beat, 8 bytes are transferred, so *Output external write beats* = 8 ∗ *Output external write bytes*. Second, a derivation happens when there exists a set of inter-dependent PMCs that describe a particular micro-architectural event, *e.g., raw-l1d-cache* is the total number access to L1 data cache and is the sum of read access given by PMC *raw-l1d-cache-rd* and write access given by PMC *raw-l1d-cache-wr*.

Finally, motivated by the need to filter PMCs for use in GPU modeling and the drawbacks of manual selection (sub-optimality, not easily transferred to a new architecture), we develop an automated PMC-based GPU power modeling methodology, APGPM. This methodology can be used to automatically derive the optimal subset of PMCs that provide the highest GPU modeling accuracy. APGPM consists of two key steps: (1) clustering the PMCs to find the inter-dependency relationship; (2) identifying their representative PMCs (for use as features in GPU modeling).

We implemented APGPM in Android and evaluated it on two representative mobile GPUs: Qualcomm's Adreno 640 (on Pixel 4) and ARM's Mali G-710 GPU (on Pixel 7) for diverse GPU workloads. Our evaluation shows that APGPM builds a GPU power model that reduces the average GPU power prediction error by almost half compared to the prior-art, utilization-frequency-based smartphone GPU power model, from 23.21% to 11.91% for Pixel 4 and from 24.72% to 9.30% for Pixel 7, across Rendering, Neural Network and Compute workloads.

We summarize our contributions as follows:

- We present a methodology for automatically selecting a minimum subset of GPU PMCs to be used as features in mobile GPU power modeling to maximize the modeling accuracy and to reduce PMC logging overhead.
- We present extensive evaluation on the two dominating mobile GPU architectures (Qualcomm's Adreno and ARM's MALI) that shows our mobile GPU power model based on APGPM reduces the average MAPE modeling error by 1.95× for Pixel 4 while using 20.41% of the total number of available PMCs and 2.66× for Pixel 7 while only using 4.66% of the total number of available PMCs, across Rendering, Neural Network and Compute workloads.

## 2 BACKGROUND ON MOBILE GPU

We provide a brief background on the two representative mobile GPUs widely used in modern smartphones. Qualcomm's Adreno series and ARM's Mali series. The two GPUs tackle the common low power and thermal dissipation constraints using different architectural designs.

***Execution Core.*** Each mobile GPU's execution pipeline has several components. Even under the same utilization, each of these pipeline components may consume different amount of power as they are associated with different types of internal structures and operations. Figure 1 shows various components of the execution core for the ARM MALI Valhall series of GPUs. The processing unit has 4 main parts: FMA (fused multiply-accumulate which is the main arithmetic computation), CVT (convert which implements simple operations like format conversion), SFU (special functions unit which implements special functions, i.e., complex functions such as computing for reciprocals and transcendental functions), and MSG (message). The execution core also houses the Varying unit, a dedicated pipeline that implements the varying interpolator, the Load/store unit which handles all non-texture memory accesses, the ZS/Blend unit which handles all accesses to the tile-memory, and the texture unit which handles all texture sampling and filtering operations. For each of these components, there are dedicated hardware

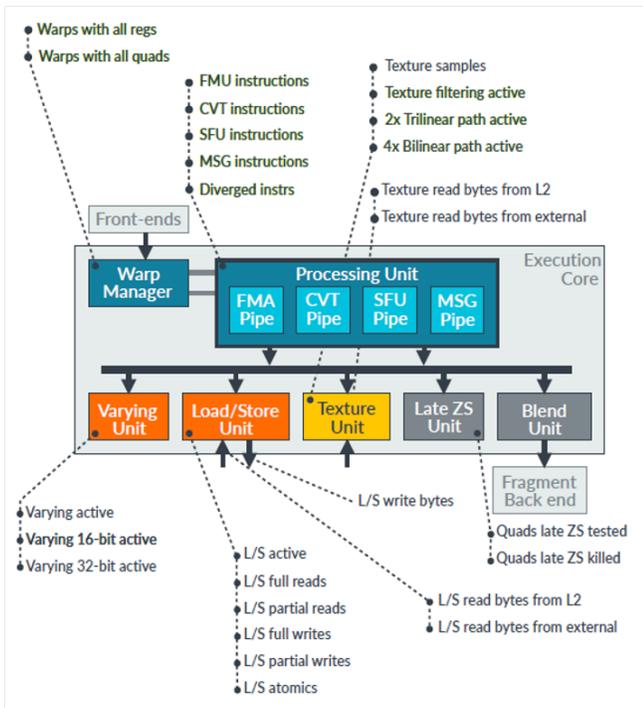

Figure 1: Execution Core schematics and PMCs for MALI-G710 [3]

Performance Monitoring Counters (PMCs) that track various events. For example, PMCs *Arithmetic FMA instructions*, *Arithmetic CVT instructions* and *Arithmetic SFU instructions* track the statistics regarding every instruction issued to the FMA, CVT and SFU, respectively.

*Memory System and Shared Memory.* The memory system is one of most power consuming component in the SoC. For example, a typical DRAM (main memory) access cost is between 80mW and 100mW per GB/s of bandwidth used. Thus, a typical memory power budget of 650 mW [4] can only sustain accessing 100 MB of the DRAM per frame at 60 FPS [3]. Also, unlike a desktop GPU which often has its own dedicated memory for the GPU and the CPU, on a mobile device the memory is shared among the GPU and CPU. The mobile phones use the memory technology known as Low Power Double Data Rate (LPDDR). Although LPDDR runs slower and has higher latency compared to DDR (used by desktops, laptops and servers), it is more power-efficient [18]. Since the memory is shared, there is memory contention between the CPU and GPU which limits the performance of memory intensive processes running simultaneously in both of them. Since the ground truth power drawn by the memory cannot be separated from the GPU (or the CPU), we will model the combined GPU and memory power by running workloads on the GPU (and similarly we model the combined CPU and memory power from running workloads on the CPU), similarly as in previous mobile CPU/GPU power modeling works [10, 25].

## 3 MOTIVATION

### 3.1 Utilization-Frequency based Model is Inaccurate

Utilization-frequency based models (e.g., [10, 25]) have been widely used for modeling GPU power on smartphones. This is due to the ease of collecting these triggers, *i.e.,* they could be easily polled from the GPU drivers. Such models perform linear regression on the GPU utilization for each GPU frequency as shown in Eq 1.

$$Power_{GPU} = \beta_{Frequency} * Utilization_{GPU} + \beta_0 \quad (1)$$

where $\beta_{Frequency}$ is the coefficient of utilization under a given GPU frequency; the coefficient typically increases with frequency as the power consumption increases with frequency. When the frequency is held constant, the GPU power draw is modelled to be linear with the GPU utilization.

To measure the accuracy of the utilization-based GPU model for modern mobile GPUs, we run several representative mobile workloads, Rendering, Neural Network and Compute (OpenCL), while keeping the GPU frequency fixed. Each workload consists of a set of benchmarks, as shown in Table 1. As the workloads are diverse in nature, we could observe different average utilization to study how utilization affects GPU power draw for diverse workloads. To study the diverse behavior, we ran the benchmarks one-by-one for each workload type with the GPU Frequency fixed at 471 MHz. Each benchmark in each workload type is repeated in a loop until reaching a fixed time duration, *e.g.,* for rendering, 60 frames were generated every second. Due to the repetitive nature of the workload, the average utilization over the duration of the run was taken. We then group the benchmarks per workload type according to the observed GPU utilization and plot the average GPU power draw per group.

Figure 2 shows the average GPU power draw under the same utilization group differ significantly for different workloads, suggesting the utilization-based power model can be highly inaccurate[1]. For example, at 70% GPU utilization, the power drawn by the Compute workload differs from the Neural Network workload by more than 30%. This happens because GPU utilization only captures the fraction of time that the GPU remains active but not the diverse microarchitecture-level usage such as diverse instructions and memory transaction breakdowns under different workloads, which contribute to different GPU power draw as shown in Figure 3 for 70% GPU utilization.

---

[2] Vision and language models were run on TFLite.
[1] Note in Figure 2, there are missing bars for some of the GPU utilization groups as we did not have benchmarks of that workloads type.

Table 1: Benchmarks and workloads used for CPU and GPU power modeling.

| Workload Type | Benchmark or workload | Num. of workloads |
|---|---|---|
| CPU | | |
| Compute | PolyBench (CPU) | 56 |
| Neural Network | TFLite[2] (CPU) | 120 |
| GPU | | |
| Compute | PolyBench (OpenCL) | 12 |
| Rendering | Vulkan | 41 |
| Neural Network | TFLite[2] (GPU) | 77 |

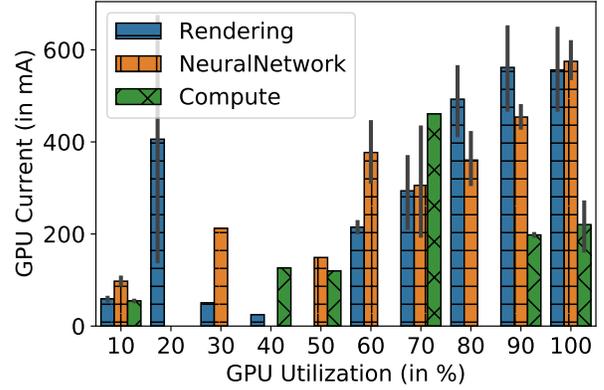

Figure 2: GPU Current vs GPU Utilization on Pixel 7 with GPU Frequency fixed at 471 MHz.

Figure 3a shows that Rendering and Neural Network workloads have higher number of instructions executed compared to Compute workloads. Even between Rendering and Neural Network workloads with similar average instructions issued per second, *i.e.,* 0.10 million, we can observe that the Rendering workload is dominated by Varying instructions, whereas the Neural Network workload is dominated by Fused Multiplication Accumulation (FMA) and Convert (CVT) instructions. This is because the Rendering workload manipulates mesh and texture which is done by the Varying unit, a dedicated fixed-function varying interpolator that uses warp vectorization for high functional unit utilization [1]. In contrast, the Neural Network workload mainly multiplies weights with tensors, which uses FMA instructions for complex math operations, and CVT instructions for simple math operations [1]. Finally, the Compute workload predominately performs simple math operations, and hence the instruction breakdown is dominated by Convert (CVT) instructions.

In terms of memory usage, as shown in Figure 3b, the memory profile for both Rendering and Neural Network workloads are similar as these have similar memory footprint of about 70 MB, whereas the Compute benchmarks being 2-D Finite Different Time Domain Kernel, have high percentages of memory transactions.

Due to the above differences among the three types of workloads and because memory instructions consume more power than arithmetic ones, we observe that at 70% GPU utilization, the Compute workload consumes on average 33.68% more GPU current than Neural Network and Rendering workloads. This is also the key reason why the GPU power for the GPU utilization groups in Figure 2 does not monotonically increase for same types of workloads. This suggests that the utilization-based power model can be highly inaccurate. We have observed similar trends for 90% and 100% GPU utilization as that for 70% GPU utilization (Figure 2).

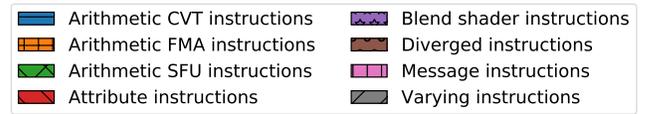

(a) Instructions breakdown

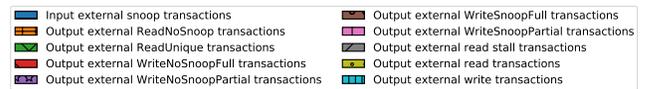

(b) Memory transactions breakdown

Figure 3: Workload breakdown on Pixel 7 with GPU Frequency fixed at 471 MHz at 70% GPU Utilization

Table 2: Number of PMCs on Modern Smartphones

| Phone | Processor Type | Processor | No. of PMCs |
|---|---|---|---|
| Pixel 4 | CPU | Little Core (A55) | 60 |
| | | Big Core (A76) | 97 |
| | GPU | Qcomm. Adreno 640 | 49 |
| Pixel 7 | CPU | Little Core (A55) | 60 |
| | | Big Core (A78) | 97 |
| | | Performance Core (X1) | 106 |
| | GPU | ARM Mali G-710 | 322 |

## 3.2 PMC-based Power Modeling for Desktop GPUs

The CPU and GPU in modern computer systems are equipped with a Performance Monitoring Unit (PMU) which contains a comprehensive set of PMCs that gather various statistics about the processor's micro-architectural events in the processor and memory system during runtime [2]. As such, they could be used as more fine-grained and potentially more suitable triggers than the simple utilization metric in GPU power modeling.

Indeed, PMC-based power modeling has been explored for desktop GPUs [11, 15, 28]. It has also been used to develop a combined CPU and GPU power model by [12]. However, all of the above PMC-based power models for desktop GPUs rely on *manually selected* CPU and GPU PMCs. As a result, they suffer the following three main limitations.

(1) Manual selection of the PMCs requires deep understanding of the GPU architecture;
(2) Manually selected PMCs may not be optimal *i.e.,* the resulting power model may provide sub-optimal model accuracy;
(3) Since PMCs are architecture specific, finding substitute PMCs for a new mobile GPU architecture may not be easy or possible. We have tried but are not able to find substitute PMCs in mobile GPUs that can represent the manually selected PMCs in the prior-art desktop/server based GPU model.

## 4 AUTOMATED PMC-BASED GPU POWER MODELING DESIGN

### 4.1 Research Questions

Given the poor accuracy of prior-art utilization-based power models for mobile GPUs, in this paper, we ask an important research question: *can PMC-based power modeling be applied to mobile GPUs, and if so, how?*

We proceed to answer this question in three steps:

**Q1: Can the PMCs of modern commercial mobile GPUs (Qualcomm's Adreno and ARM's MALI) be used to model the GPU power?** Since mobile devices are power-constrained, the mobile GPU manufacturers may be highly judicious in adding hardware PMCs to the architecture as they incur extra overheads such as silicon die area, design complexity as they need to be read out by the CPU, latency of execution pipelines, and ultimately the power efficiency of the GPU. As such, it is unclear whether the PMCs available on mobile GPUs are comprehensive enough in capturing all architectural components of the GPU and hence the overall power draw of the whole GPU.

**Q2: Are the PMCs in mobile GPUs inter-dependent and hence needing to be filtered when used in power modeling in order to maximize the modeling accuracy and reduce PMC logging overhead?** If the answer to **Q1** is yes, there may be too many PMCs available on mobile GPUs as shown in Table 2 such that some of them are *inter-dependent*. Since including inter-dependent variables in GPU power modeling, *e.g.,* in linear regression, can result in lower modeling accuracy and hence be counter-productive, we need to filter such inter-dependent PMCs.

**Q3: For a given mobile GPU, how to automatically filter PMCs down to a subset of PMCs for use in accurate PMC-based power modeling?** If the answer to **Q2** is yes, filtering inter-dependent PMCs needs to be automated and efficient as either manual filtering or a compute-intensive filtering process would hinder the practical use of the PMC-based mobile GPU power modeling methodology.

### 4.2 Methodology: How to Isolate GPU Power?

Studying the correlation between GPU PMCs and GPU power draw requires a way to accurately measure the ground truth GPU power draw during the execution of a benchmark workload. However, on most phones we can directly measure only the total power draw by attaching an external power monitor, *e.g.,* a Monsoon power monitor, and not the power draw by the GPU housed inside the SoC.

***Isolating GPU power.*** Our methodology for isolating the GPU power from the total power measured hinges on two observations: (1) When we perform GPU experiments, we can turn off all other components like display except for CPU and base power. (2) When we perform CPU experiments, we can even turn off the GPU. Based on these two observations, we first build a power model to estimate the CPU power and then we subtract the CPU and Base power from the total power measured by the power monitor in GPU experiments to obtain the ground truth GPU Power.

*(1) Base Power.* The phone Base power comprises the constant static power drawn by the entire SoC which includes both the CPU and GPU. To get the base power, we use a dummy app that turns the phone display off and acquires a wake-lock to keep the CPU on and sleeps (hence incurring no workload). During the experiment, only background system processes are running occasionally which incurs insignificant CPU utilization and hence CPU power draw. The total power measured by the Monsoon power monitor is considered as the Base power.

*(2) CPU Power.* To estimate the CPU power, we utilize a methodology that mirrors that for the GPU: (1) Run CPU benchmarks, (2) Collect CPU PMC values and ground-truth CPU power (*i.e.,* the total power minus the base power); (3) Identify statistically significant PMC clusters and apply linear regression on the representative PMCs of those clusters to derive the CPU power model.

Table 1 shows the two types of workloads used to model the CPU power, Compute and Neural Network. For Compute, we use 56 workloads from PolyBench [8], a well-known CPU benchmark. For Neural Network, we use 120 vision and language based TFLite models [19] that run on the CPU.

**Benchmarks for GPU power modeling.** For GPU power modeling, we use three types of GPU specific workloads as shown in Table 1. For the Compute workload, we use 12 workloads from PolyBench [8], a well-known OpenCL benchmark. For Rendering, we modified 41 Vulkan examples [9] [23] to make them run with the phone display turned off. Lastly, for Neural Network, we run 77 vision and language based TFLite models [19] that run on the GPU. Our proposed PMC-based power model is trained using these representative set of workloads which exercise most components of a mobile GPU and thus is expected to accurately model the GPU power for any application run. A few application case studies are presented in Section 6.

## 4.3 Are PMCs sufficient to Model Mobile GPU Power?

We first measure the correlation of individual PMCs of two representative mobile GPUs with the GPU power draw using the GPU workloads discussed above. In particular, we calculate the Pearson correlation between the PMC's event rate and the corresponding GPU power. Figure 4 shows the sorted correlation between the GPU PMCs and GPU power draw for Pixel 4 and Pixel 7. It shows that 27 out of the total 49 PMCs for Pixel 4 and 221 out of the total 322 PMCs for Pixel 7 are positively correlated with the GPU power; the Pearson correlation is in the range of 0.10 to 0.61 on Pixel 4 and 0.10 to 0.67 on Pixel 7. We consider a correlation as statistically significant if it has a *p*-value less than 0.05 as shown in the figure by a horizontal red line. Event rates of 1 PMC on Pixel 4 and 24 PMCs on Pixel 7 are always captured as zero; we consider them as having no correlation with the GPU power. The above results show that several available PMCs are positively correlated with the GPU power draw,

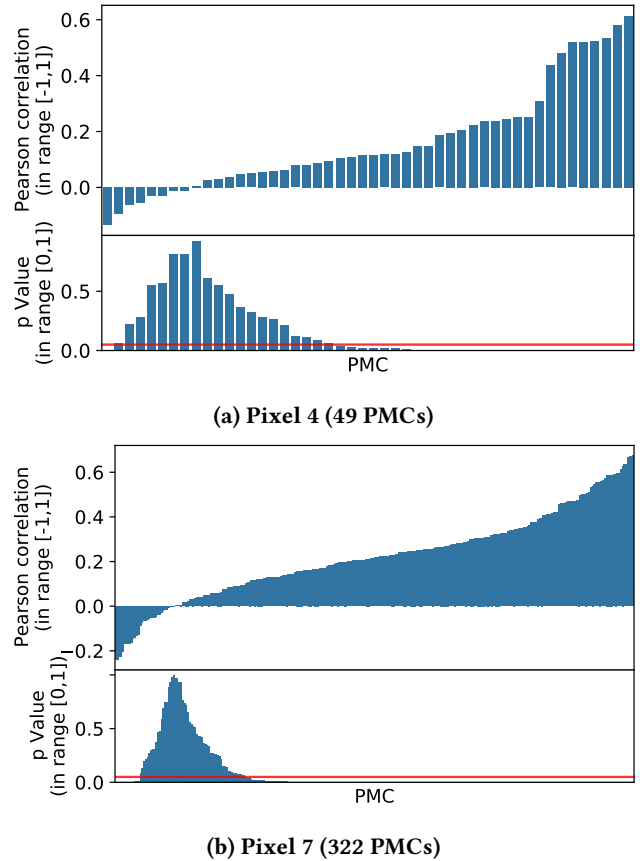

(a) Pixel 4 (49 PMCs)

(b) Pixel 7 (322 PMCs)

**Figure 4: Pearson correlation between GPU PMC event rates and the GPU Power for Pixel 4 and Pixel 7.**

suggesting that together they may be sufficient to model the GPU power draw.

Next, we directly answer **Q1** – whether the set of available PMCs are sufficient in modeling the GPU power draw, by measuring the *coefficient of determination* or $R^2$ [7] when performing multivariable linear regression between event rate of all the PMCs (independent variables) with the GPU power (dependent variable); a value of the coefficient of determination or $R^2$ close to 1 signifies that the independent variables can indeed explain the dependent variable. We find that the $R^2$ values are 0.93 for Pixel 4 and 0.99 for Pixel 7, suggesting that the available PMCs on these GPUs are sufficient to model the GPU power.

**Some PMCs have negative correlation with GPU power.** Figure 4 also shows that some of the PMC events are negatively correlated with the GPU power. For each such PMC, we invert the PMC value to create the *inverted PMC* which will then have a positive correlation with the GPU power. As examples of inverted PMCs, we found that some PMCs on mobile GPUs count stalling of compute pipelines or memory transactions. During such events, the system has to wait for acquiring the resources so that the corresponding pipeline can resume. The inverted PMC event rate effectively captures how the stalls negatively affect the GPU power. In total, we found 1 inverted PMC on Pixel 4 and 14 inverted PMCs on Pixel 7.

## 4.4 Are Mobile GPU PMCs Inter-dependent?

Building on the positive answer to **Q1**, we next study **Q2** – whether the PMCs are inter-dependent. If PMCs are inter-dependent, they need to be filtered when used as features in the GPU power model, because multicollinearity among independent variables reduces statistical model's regression power [6].

***Statistical interdependence of GPU PMCs.*** Multicollinearity occurs when the independent variables are correlated with each other. In Figure 5, we show the PMC-PMC Pearson correlation as a heat map for 10 PMCs that have individually the highest correlation with the GPU power draw. We see that many pairs of PMCs (*e.g.,* shaded in dark blue) are highly correlated; these PMC pairs are statistically interdependent. This happens for two important reasons: (1) Some PMCs are related to each other by conversion, *e.g.,* in Pixel 7, *Output external write beats* = 8 ∗ *Output external write bytes*; (2) Some PMCs are related by derivation. This happens as there may exist a set of inter-dependent PMCs that describe a particular micro-architectural event. For example, in Pixel 4's CPU, which is ARM Cortex A76 (Big core), PMC *raw-l1d-cache* is the total number access to L1 data cache and is the sum of read access given by PMC *raw-l1d-cache-rd* and write access given by PMC *raw-l1d-cache-wr*, so *raw-l1d-cache = raw-l1d-cache-rd + raw-l1d-cache-wr*

***Combined GPU PMCs may have higher correlation.*** We also examine whether combining two or more PMCs, *i.e.,* using their product value, can have higher correlation with the GPU power. Table 3 shows some of such combined terms from combining two PMC events (*i.e.,* their product) indeed have higher correlation with the GPU power compared to the individual PMC events. This happens when the combined PMCs can better capture the mobile GPU's sub-component usage at the micro-architecture level than individual PMCs. For example, on Pixel 7, power consumed by external write i.e., from L2 cache to external memory can be explained by the number of write transactions to external memory. One of the ways to derive it could be from the product of *L2 cache write miss ratio* (*i.e.,* Percentage of L2 cache misses which result in external writes) and *GPU active cycles* [3].

## 4.5 APGPM: Automated PMC-based GPU Power Modeling

Our findings above suggest that the PMCs in mobile GPUs are sufficient to model the GPU power, but need to be filtered into a subset of PMCs to remove multicollinearity which negatively affects power modeling accuracy [6]. Manually selecting a subset of PMCs would suffer the same limitations as in previous work for desktop GPU power modeling (§3.2): (1) it requires expert understanding of the GPU architecture; (2) manually selected PMCs may not be optimal; and (3) the selected PMCs cannot be reused in a new architecture.

In this paper, we develop an automated PMC-based GPU power modeling methodology, APGPM, that can be used to automatically derive the optimal subset of PMCs that provide the highest GPU power modeling accuracy. APGPM consists of two main steps:

(1) Cluster the PMCs to find the inter-dependency relationship;
(2) Form significant PMC clusters and identify their representative PMCs.

*4.5.1* ***Step 1: Clustering.*** First, we normalize the PMC event rates, *i.e.,* (x - mean of x) / (standard derivation of x). Then, we apply Agglomerative clustering with Ward linkage [22, 27]. We choose ward linkage as it is shown to give the merged clusters with the least variance and could handle outliers. We tried various distance thresholds and found 0.05 times the number of samples to be the most efficient in clustering the PMCs.

*4.5.2* ***Step 2: Significant Clusters and Representative PMCs.*** There are several ways to select the features to be used in a statistical regression model: (1) *forward stepwise selection* adds independent variables one by one until adding additional ones does not improve the model to a statistically significant extent; (2) *backward stepwise elimination* starts with all the independent variables in the system and removes them one by one until we reach the best $R^2$; (3) *hybrid bidirectional elimination method* uses both forward and backward passes in turn. Additionally, we can add combination of independent variables as features as described in §4.4. However, using combined PMCs can be very expensive as the number of combinations increases polynomially with the number of variables.

In APGPM, we use a hybrid feature selection approach: (1) it first finds significant PMC clusters and their representative PMCs. (2) it then determines which significant PMC clusters to use in GPU power modeling.

***Identifying the representative PMC per cluster.*** First, we apply single variable linear regression between each of the cluster's elemental PMCs one by one with the GPU power and calculate their $R^2$. We denote the maximum $R^2$ among the elemental PMCs as the *cluster importance*, and the PMC associated with the cluster importance is called the *representative PMC* of the cluster.

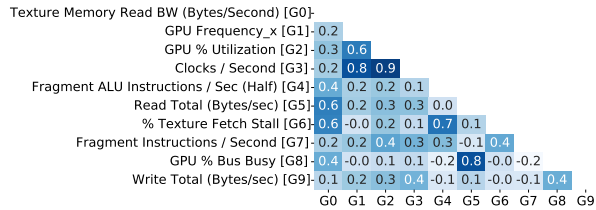
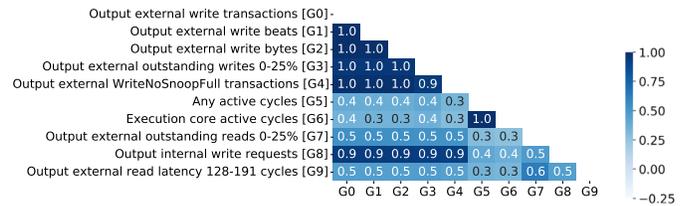

(a) Pixel 4     (b) Pixel 7

Figure 5: Pearson correlation between GPU PMC vs PMC event rate for Pixel 4 and Pixel 7 for the 10 PMCs that have the highest correlation with GPU power.

Table 3: The Pearson correlation between the GPU power with 2 individual PMCs and with their combined PMC term.

| PMC1 | Corr. PMC1 | PMC2 | Corr. PMC2 | Combined Corr. |
|---|---|---|---|---|
| | | Pixel 4 | | |
| Fragment Instructions / Second | 0.43 | Read Total (Bytes/sec) | 0.51 | 0.75 |
| Fragment Instructions / Second | 0.43 | GPU % Bus Busy | 0.30 | 0.71 |
| GPU Frequency | 0.58 | Texture Memory Read BW (Bytes/Second) | 0.61 | 0.67 |
| | | Pixel 7 | | |
| Any iterator active cycles | 0.55 | L2 cache write miss rate | 0.49 | 0.77 |
| Message instructions | 0.47 | Output external ReadNoSnoop transactions | 0.58 | 0.76 |
| GPU active cycles | 0.53 | L2 cache write miss rate | 0.49 | 0.76 |

***Identifying significant clusters.*** Next, instead of simply using the representative PMCs, one each from the clusters, as the features in GPU power modeling, we search through the clusters to choose the minimal subset of *significant clusters* whose representative PMC together will best model the GPU power. Using the fewest PMCs in the model minimizes the logging overhead, *i.e.*, from reading the PMCs at runtime.

First, we sort the clusters in the descending order of their cluster importance. This order approximates the sorted contribution of the clusters when their representative PMCs are used in modeling the GPU power. We then go through the list of sorted clusters one by one to decide if each cluster should be included in the significant clusters (SC) set, as follows.

(1) We select the cluster having the highest cluster importance, *i.e.*, with its representative PMC having the maximum $R^2$ as the first cluster in the SC set. We assign its $R^2$ value to the current $R^2$ value of the SC set denoted as $R^2_{sc}$.
(2) To check whether the next cluster $C_i$ in the sorted list can be added to the SC set, we apply multi variable linear regression between each PMC of $C_i$ and the representative PMCs of the current clusters in the SC set (as independent variables) with the GPU power and calculate the total $R^2$ value.
(3) The representative PMC for candidate cluster $C_i$ is the one that gives the maximum total $R^2$ when including $C_i$.
(4) We compare the above maximum $R^2$ when including $C_i$ with $R^2_{sc}$. If $R^2_{sc}$ is less than maximum $R^2$, we add $C_i$ to the set of significant clusters and update $R^2_{sc}$ value with maximum $R^2$.

The above steps are repeated for the remaining clusters until $R^2_{sc}$ values stop growing beyond a threshold, *i.e.*, $R^2_{sc}$ has not increased beyond a threshold in the last five iteration, we terminate the search. We use a threshold of 0.01 for our termination condition.

## 5 EVALUATION

### 5.1 Methodology

We perform evaluation on representative mobile GPUs from two dominating mobile GPU families widely used in modern smartphones: ARM's Mali G-710 GPU on Pixel 7 and Qualcomm's Adreno 640 GPU on Pixel 4. We use a Monsoon power monitor for measuring the total phone power. To reduce measurement noise, we kept the setup thermally cooled by applying cold packs and used the internal temperature sensor to programmatically start the experiments only when the devices are sufficiently cold, *i.e.,* when the temperature is below 40°C. Additionally, to minimize the amount of bloatware, we used the engineering build of AOSP, and uninstalled or disabled any other apps that were included in the engineering build of AOSP. We also kept the phone on airplane mode with the display off. We ran all the workloads in Table 1 at three different GPU frequencies; 257 MHz, 345 MHz and 427 MHz for Pixel 4; and 251 MHz, 351 MHz and 471 MHz for Pixel 7. We split the entire runs into 2/3 for training and 1/3 for testing, and we have used this split for all the models evaluated.

Table 4: Clustering for CPU power modeling.

|  | Pixel 4 | Pixel 7 |
|---|---|---|
| Number of CPU PMCs | 92 | 106 |
| Number of clusters | 11 | 10 |
| Number of significant clusters | 6 | 6 |

Table 5: Accuracy for CPU power modeling. Error shown are shown as mean (median).

| Phone |  | $R^2$ | MAE (in mA) | MAPE (in %) |
|---|---|---|---|---|
| Pixel 4 | Train | 0.87 | 21.53 (19.53) | 8.11 (6.78) |
|  | Test | 0.88 | 19.10 (14.87) | 6.63 (5.40) |
| Pixel 7 | Train | 0.86 | 23.91 (17.17) | 6.15 (3.76) |
|  | Test | 0.82 | 24.50 (19.90) | 5.73 (4.48) |

## 5.2 CPU Power Modeling

Table 4 shows the clustering result for PMC-based CPU power modeling. For Pixel 4, we generated a power model for Cortex A76 and for Pixel 7, we generated a model for Cortex X1. We started with 90+ PMCs in both cases and extracted 6 significant clusters. Table 5 shows the CPU power models using the representative PMCs from these significant clusters achieve 6.63% and 5.73% mean absolute percentage error for Pixel 4 and Pixel 7, respectively, which translate to 19.10 mA and 24.50 mA mean absolute error.

## 5.3 GPU Power Modeling

First, we evaluate a linear model using only the base PMCs (*i.e.,* not using inverted or combined PMCs). We denote this model as Linear APGPM. Table 6 shows the clustering process for GPU power modeling. APGPM started with 49 PMCs for Adreno 640 and 322 PMCs for MALI G-710 and clustered them into 11 and 13 significant clusters, respectively. Table 7 shows that the GPU power models using the representative PMCs from these significant clusters achieve 13.03% and 10.76% mean absolute percentage error for the two GPUs, respectively, which translate into 40.34 mA and 39.26 mA mean absolute error.

Table 6: Clustering for GPU power modeling.

|  | Pixel 4 | Pixel 7 |
|---|---|---|
| Number of GPU PMCs | 49 | 322 |
| Number of clusters | 18 | 80 |
| Number of significant clusters | 11 | 13 |

Table 7: Accuracy for GPU power modeling only using linear PMCs. (Error shown are shown as mean (median).)

| Phone |  | $R^2$ | MAE (in mA) | MAPE (in %) |
|---|---|---|---|---|
| Pixel 4 | Train | 0.87 | 35.02 ( 25.81) | 11.78 ( 7.33) |
|  | Test | 0.76 | 40.34 ( 28.11) | 13.03 ( 8.53) |
| Pixel 7 | Train | 0.90 | 32.34 ( 23.27) | 10.56 ( 6.93) |
|  | Test | 0.72 | 39.26 ( 19.79) | 10.76 ( 5.95) |

Table 8: Number of PMCs used by APGPM.

| Phone | Model | Number of PMCs used | Percentage of PMCs used (in %) |
|---|---|---|---|
| Pixel 4 | Linear APGPM | 11 | 22.45 |
|  | APGPM | 10 | 20.41 |
| Pixel 7 | Linear APGPM | 13 | 4.04 |
|  | APGPM | 15 | 4.66 |

Importantly, Table 9 shows 6 sample PMC clusters out of the 13 significant clusters on Pixel 7. We observe that different clusters correspond to disjoint parts of the GPU micro-architecture which all contribute to the total GPU power consumption. This explains why a linear regression model using the PMCs is sufficient and can be accurate in modeling the total GPU power.

Next, we evaluate the impact of including inverted (§ 4.3) and combined PMCs, *e.g., GPU active cycles* and *L2 cache write miss rate* can be combined to create a new combined PMC (§ 4.4). We denote this model as APGPM. Here, we applied APGPM to combined PMCs to create clusters and generated a linear regression model by considering the representative combined PMCs from the significant clusters. In particular, we have only considered pairwise combination of all available PMCs, *i.e.,* their product and division, to limit the search space. Table 10 shows that including such combined PMCs reduces the absolute percentage error by 1.12% (to 11.91%) for Pixel 4 but increases the error by 1.46% (to 9.30%) for Pixel 7. Table 8 shows that APGPM only used 20.41% and 4.66% of the total number of available PMCs for Pixel 4 and Pixel 7, respectively. This shows huge reduction in the logging overhead.

**Table 9: Some of the selected significant clusters and their representative PMCs.**

| PMCs in the significant clusters | Representative PMC |
|---|---|
| Execution pipeline | |
| Arithmetic FMA instructions Arithmetic instruction issues | Arithmetic instruction issues |
| Arithmetic CVT instructions Message instructions | Message instructions |
| Texture filtering cycles Texture message read beats Texture message write beats | Texture filtering cycles |
| Memory system | |
| Output external ReadNoSnoop transactions Output external outstanding reads 0-25% Output external read beats | Output external outstanding reads 0-25% |
| Input internal read stall cycles Output internal read stall cycles Texture fetch stalls | Texture fetch stalls |
| L2 cache write miss rate | L2 cache write miss rate |

**Table 10: Accuracy for GPU power modeling using inverted and combined PMCs. (Error shown are shown as mean (median).)**

| Phone | | $R^2$ | MAE (in mA) | MAPE (in %) |
|---|---|---|---|---|
| Pixel 4 | Train | 0.88 | 32.09 ( 24.65) | 10.94 ( 7.04) |
| | Test | 0.83 | 34.80 ( 25.05) | 11.91 ( 6.99) |
| Pixel 7 | Train | 0.92 | 30.48 ( 22.95) | 9.59 ( 6.60) |
| | Test | 0.81 | 34.87 ( 22.77) | 9.30 ( 6.33) |

### 5.4 APGPM selects PMCs that capture GPU Micro-architecture

We mapped the significant clusters of PMCs shown in Table 9 to various parts of the micro-architecture for MALI G-710 GPU on Pixel 7 as shown in Figure 6. Figure 6a shows how the instruction-based PMCs are mapped to various parts on the main execution unit. The First cluster has PMCs that captures events in the FMA unit which is the main computation unit. The second cluster captures events in the CVT and message processing units which are responsible for simple math operations. The third cluster captures events in the texture processing unit required for rendering. For memory transactions, the GPU has three main units on the memory system which are the internal read and write ports which connect the L2 cache with the GPU, the L2 cache, and the the external read and write ports which connect the main memory to the L2 cache. Figure 6b shows that the three clusters selected by the APGPM indeed captures events in these three sub-systems. These results suggest that APGPM can

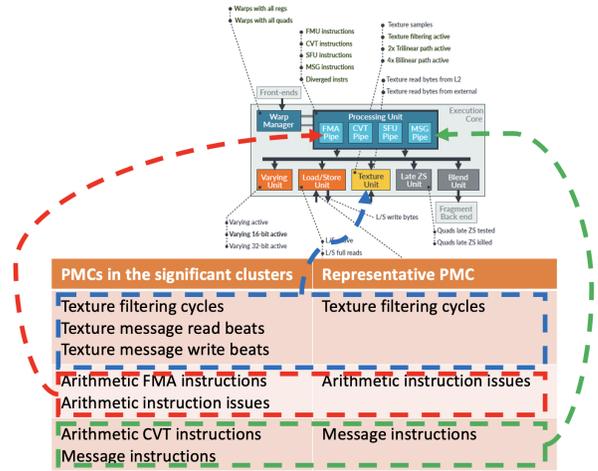

(a) Execution pipeline PMCs cluster in Table 9 mapped to execution core parts on MALI G-710 GPU. (Execute core schematic taken from [3])

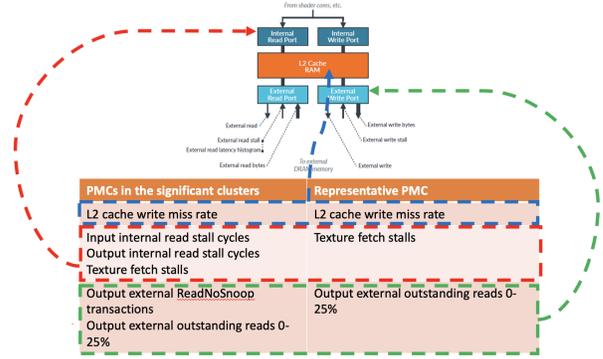

(b) Memory system PMCs cluster in Table 9 mapped to execution core parts on MALI G-710 GPU. (Memory core schematic taken from [3])

**Figure 6: APGPM automatically selects PMCs which the describes micro-architecture**

automatically select clusters of PMCs that capture all the important events in the GPU micro-architecture.

### 5.5 Comparison with Baseline Models

Next, we compare APGPM with four baseline models: (1) the prior-art utilization-frequency model; (2) *All-PMC*, a linear regression (LR) model using all the available PMCs; (3) *k-top-PMC* [14], which uses *i.e.,* linear regression (LR) with the *K* PMCs that have the highest Pearson correlation with the GPU power – the model uses the same number of PMCs as that of APGPM, *i.e.,* 10 for Pixel 4 and 15 for Pixel 7; and (4) *NN-All-PMC:* a multilayer perceptron model with 3 fully-connected layers using all PMCs as input features, to
10

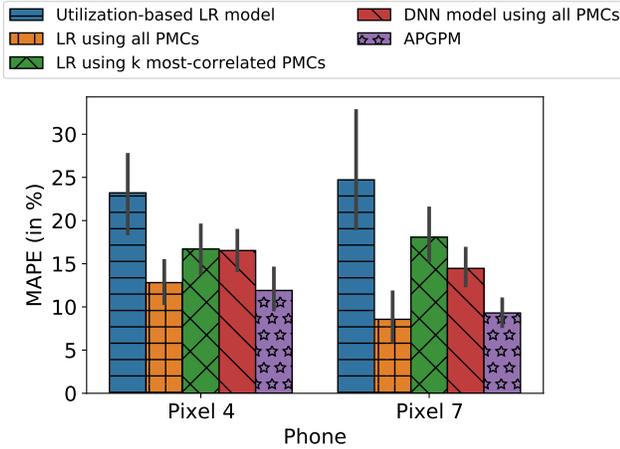

Figure 7: Comparison of APGPM model with baseline models (k is taken as 10 for Pixel 4 and 15 for Pixel 7 which are corresponding to the number of PMCs used by APGPM).

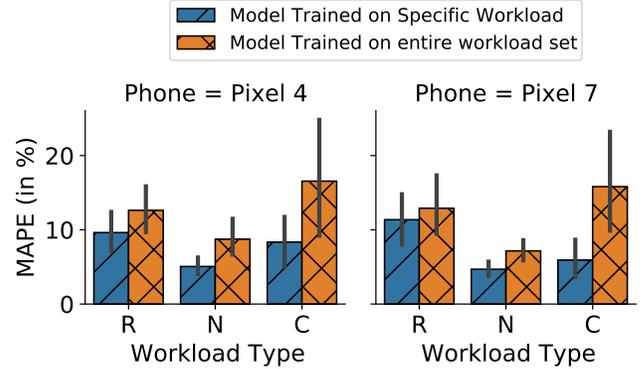

Figure 8: Impact on benchmark workloads by customizing PMC selection (R: Rendering, N: Neural Network and C: Compute).

evaluate whether a DNN-based model can automatically pick the important PMCs and achieve high modeling accuracy.

Figure 7 shows the comparison results and we make the following observations, First, compared with the utilization based models, APGPM reduces the average MAPE error from 23.21% to 11.91% (by 1.95×) for Pixel 4 and from 24.72% to 9.30% (by 2.66×) for Pixel 7. Second, compared with All-PMC, the average MAPE for APGPM differs by 0.91% for Pixel 4 while using 20.41% of the total PMCs and 0.73% for Pixel 7 while using 4.66% of the total number of available PMCs. This shows that APGPM can have similar accuracy with a small effective set of PMCs while hugely reducing the logging overhead.

Third, compared to k-top-PMCs, APGPM-generated model reduces the MAPE by 4.79% on Pixel 4 while using 10 PMCs and by 8.78% on Pixel 7 while using 15 PMCs. These results suggest that APGPM is able to choose a minimal set of representative PMCs while achieving the highest modeling accuracy.

Finally, Figure 7 also shows that the NN-based power modeling using all PMCs achieves lower accuracy than APGPM; compared to APGPM, the MAPE is 4.62% and 5.17% higher on Pixel 4 and Pixel 7, respectively. These results suggest that the multilayer perceptron model cannot effectively identify the more important PMCs for use in GPU power draw prediction.

## 5.6 PMC selection for specific workloads

Next, we study the impact of benchmark workload used in APGPM on the resulting model accuracy. Figure 8 shows that the accuracy of the models created with PMCs selected based on training on specific workload types, denoted as specialized models, differs from that created with PMCs selected using the entire workload (of all three types), denoted as the general model. In particular, on Pixel 4 and Pixel 7 respectively, the MAPE of the general model is 3.00% and 1.54% higher than that of the model trained for the Rendering workload when tested on the Rendering workload, and 3.68% and 2.47% higher than the model trained for the Neural Network workload when tested on the Neural Network workload.

For Compute workload, the error of the general model is much worse than specialized models, *i.e.*, by 8.21% and 9.90% on Pixel 4 and Pixel 7, respectively. This is likely due to the limited number of PolyBench benchmarks available.

## 5.7 Overhead Analysis

On a mobile device, GPU power model requires logging raw PMC events and post process the raw PMC event trace to get the required average PMC event rate as the input to the model. It is important to keep both overheads to a minimum.

*5.7.1 Power overhead for logging raw PMC events.* Table 11 compares the overhead of APGPM with the GPU model using all the PMCs. We observe APGPM consumes on average 12.89% and 39.10% less power compared to logging all PMCs on Pixel 4 and Pixel 7, respectively. Since the model profiles the app load for the complete app run duration, using a small effective set of PMCs is helpful in extending the battery life without compromising the model accuracy which differs by only 0.91% for Pixel 4 and 0.73% for Pixel 7 as shown in § 5.5.

**Table 11: Overhead comparison between using all PMCs and selected PMCs by APGPM.**

| Phone | Workload Type | Overhead for logging raw PMC event rates (in mA) | | Post-processing time for raw PMC event trace (in ms) | |
| --- | --- | --- | --- | --- | --- |
| | | All PMCs | APGPM | All PMCs | APGPM |
| Pixel 4 | Rendering | 15.92 | 8.41 | 166.70 | 38.39 |
| | Neural Network | 27.02 | 26.47 | 157.89 | 37.94 |
| | Compute | 12.79 | 10.40 | 155.62 | 36.28 |
| | Combined | 22.12 | 19.26 | 160.30 | 37.89 |
| Pixel 7 | Rendering | 50.11 | 24.40 | 491.06 | 24.65 |
| | Neural Network | 77.11 | 51.94 | 485.03 | 23.96 |
| | Compute | 69.14 | 32.11 | 537.70 | 22.73 |
| | Combined | 68.08 | 41.46 | 492.84 | 24.03 |

*5.7.2 Time required to process raw PMC event trace to get the average PMC event rate.* GPU PMC values reset after a small duration, so the raw GPU PMC events have to be captured several times during the app run and these values are to be dumped from memory buffer to mobile disk periodically. Further, we need to process the raw PMC event rate trace (dumped in the mobile disk) to get the average PMC event rate which is the input to the power model. Table 11 also shows the average time required for post processing 1-second event trace for an app. For Pixel 7 with ARM Mali G-710 GPU, it takes on average 492.84 ms to process data collected for all 322 PMCs compared to 24.03 ms to process data collected for 15 PMCs of APGPM, a 95.12% reduction. Similarly, for Pixel 4 with Qualcomm's Adreno 640 GPU, it takes on average 160.30 ms to process data collected for all 49 PMCs compared to 37.89 ms to process data collected for 10 PMCs of APGPM, a 76.36% reduction.

*5.7.3 Changes in the application behavior due to PMCs logging.* From the case study in Section 6, we observed no noticeable changes in the application behavior. We observed only a small 0.66% and 1.61% increase in the average inference latency for the Neural Network workloads while logging 10 and 15 PMCs by APGPM for Pixel 4 and Pixel 7, respectively. This shows that the PMC logging has negligible effect both on the application behavior and the power modeling accuracy.

## 6 CASE STUDIES

In this section, we present two case studies showing how app developers can make use of accurate GPU power models generated by APGPM.

### 6.1 Finding Optimal Operating Point for Neural Network

For our first case study, we show how the GPU power model can help developers find the optimal operating point of deep neural networks (DNNs) for an analytic task, by trading off model accuracy with power consumption.

Consider a developer needs to choose from 5 variations of EfficientNet-Lite, a popular image classification DNN, to run on Pixel 7 to satisfy a certain app performance requirement under a GPU power constraint. Table 12 shows these 5 variations with their inference latencies. We ran these models at three different GPU frequencies, 251 MHz (lower range), 471 MHz (mid range) and 701 MHz (upper range). We could not run the model at 848 MHz and 771 MHz, *i.e.,* the two highest frequencies, as the phone thermally throttles within seconds. We tabulated the GPU current estimated by our power model, inference time and GPU energy per inference for these 5 variants at each of the three frequencies. We compute GPU energy per inference (in mWs) as the product of GPU power consumed during inference (in mW) and inference time (in seconds).

The developer can then find the optimal operational point using this table 12. For example, if the developer needs to run DNN inference at 30 FPS, *i.e.,* the inference needs to finish within 33.3 ms as highlighted in the table 12. The table shows that lite 2 and lite 4 models while running at 251 MHz and 701 MHz, respectively, can meet the 33.3 ms latency criteria. The developer can choose to run lite 2 model at 251 MHz to save 3.0 times energy per inference by sacrificing 3.2% accuracy.

### 6.2 Do AR Tasks Need Offloading? A Power Prospective

For our second case study, we show how APGPM can accurately inform AR developers whether an SoC power budget is exhausted, thus giving them a tool to decide whether to run an AR task locally on the mobile GPU or whether offloading (*e.g.,* [5]) is necessary.

Augmented Reality (AR) apps need to perform analytics (*e.g.,* using DNNs) on each camera frame which may consume much GPU power on the mobile device. An alternative to support such analytics is to offload DNN inferences to edge GPU servers. However, offloading DNN inference requires uploading large camera frames which may consume significant energy in using the wireless network interface (Wi-Fi or cellular). From a power perspective, it is unclear which approach may consume more battery in the mobile device.

The AR application developer can answer this question by using accurate power models for the mobile GPU and for the wireless NIC. We measured the mobile GPU power draw for 10 DNN-based AR analytics tasks using the APGPM-generated GPU model at 471 MHz (mid range frequency) on Pixel 7. Table 13 shows the estimated GPU current for common AR tasks. We observe that two of the most critical AR tasks, depth occlusion and object detection, together consume about 904.52 mA which is approximately 3.43 W. This would exhaust the 3W power budget as specified by ARM [4] and drains the phone battery quickly even without considering other parts of the SoC.

Table 12: Inference latency and GPU energy per inference for EfficientNet-Lite image classifiers while varying GPU frequencies on Pixel 7. (Latencies highlighted satisfies the inference timing constraint of 33.3 ms *i.e.*, for 30 FPS)

| | | | GPU Frequency | | | | | | | | |
|---|---|---|---|---|---|---|---|---|---|---|---|
| | | | 251 MHz | | | 471 MHz | | | 701 MHz | | |
| Model Type | Inference Accuracy | MAC | GPU Current (in mA) | Inference Latency (in ms) | GPU Energy per inference (in mWS) | GPU Current (in mA) | Inference Latency (in ms) | GPU Energy per inference (in mWS) | GPU Current (in mA) | Inference Latency (in ms) | GPU Energy per inference (in mWS) |
| Lite 0 | 74.1 | 407 | 242.39 | 14.81 | 13.86 | 454.60 | 8.45 | 14.82 | 659.11 | 6.35 | 16.15 |
| Lite 1 | 75.9 | 631 | 257.81 | 20.17 | 20.08 | 495.86 | 11.45 | 21.92 | 694.50 | 8.62 | 23.12 |
| Lite 2 | 77.0 | 899 | 295.06 | 27.11 | 30.88 | 553.73 | 15.24 | 32.58 | 770.68 | 11.54 | 34.34 |
| Lite 3 | 79.0 | 1440 | 317.73 | 38.21 | 46.87 | 596.36 | 20.63 | 47.48 | 834.22 | 16.42 | 52.88 |
| Lite 4 | 80.2 | 2640 | 330.52 | 61.54 | 78.51 | 626.42 | 35.19 | 85.08 | 889.51 | 26.99 | 92.68 |

Table 13: Do AR tasks need to be offloaded? From a power perspective.

| AR Tasks | GPU Current (in mA) |
|---|---|
| Depth Occlusion | 264.14 |
| Image Segmentation | 277.45 |
| Face Detection | 173.58 |
| Face Tracking | 296.21 |
| Image Classification | 320.80 |
| Pose Tracking | 382.09 |
| Face Landmark | 222.40 |
| Gesture Recognition | 284.38 |
| Hand Pose Detection | 324.11 |
| Object Detection | 640.38 |

## 7 RELATED WORK

***Utilization-based GPU power modelling.*** Utilization-frequency models are widely used because of their simplicity [13, 17, 20, 24, 26]. Such a model predicts GPU power using linear regression on GPU utilization and frequency. [10] proposed utilization based GPU power model for Nexus 4 with Adreno 320. [25] presented a GPU power model for Exynos-4412 SoC equipped with ARM Mali-400. [11] presented a statistical regression model for a NVidia GeForce 8800 GT graphics card by analyzing and modeling GPU power consumption.

***PMC-based GPU power modelling.*** [15] presented a linear regression model to estimate the average power draw of GeForce 285 GTX GPU using its PMCs. [28] proposed a GPU power model for an ATI Radeon HD5870 GPU, by correlating the GPU power draw with the architectural behaviors. [12] presented a DNN CPU-GPU power model based on manually selected 14 GPU PMCs and 12 CPU PMCs; the DNN-based model took several days to train. [14] presented a CPU and GPU power modeling by technique selecting k-highly correlated PMCs with power.

## 8 CONCLUSION

In this work, we showed that the widely used utilization-frequency power model have poor accuracy as they inherently cannot capture the diverse micro-architectural usage of modern mobile GPUs. We presented the automated power modeling methodology that automatically derives the optimal set of GPU PMCs to be used as features in linear-regression based power model to accurately estimate mobile GPU power draw. Our PMC-based mobile GPU power model reduces the MAPE modeling error of prior-art utilization-based power model 1.95× to 2.66× on two representative mobile GPUs, Qualcomm's Adreno 640 (on Pixel 4) and ARM's Mali G-710 GPU (on Pixel 7) while using only 4.66% to 20.41% of the total number of available PMCs.

We further presented two use cases of our accurate mobile GPU power model: how an app developer can choose the optimal operating point of DNNs to maintain a delicate balance between accuracy and the GPU battery drain, and how an AR developer can decide whether running an AR analytics task locally on the mobile GPU or offloading to an edge server over the wireless network will meet a given power budget.